\documentstyle[a4,11pt,epsfig,twoside]{article}
\begin{document}
\setcounter{page}{1}
\title{$t\bar{t}$ CROSS-SECTION AND FORWARD-BACKWARD ASYMMETRY AT CLIC
} 
\author{L. SALMI$^1$ \thanks{e-mail address: Laura.Salmi@cern.ch}
\\
\\
        $^1$ {\it Helsinki Institute of Physics, Helsinki}
}
\date{}
\maketitle
\begin{abstract}
A top-quark reconstruction method at {\sc Clic} is described.  The reconstructed events are used to measure the $e^+ e^- \rightarrow t \bar{t}$ cross-section as well as the forward-backward asymmetry.  The results suggest that the measurement accuracy for the cross-section will be $\frac{\delta \sigma} {\sigma}=1.4$~\% and for the asymmetry $\frac{\delta A_{FB}}{A_{FB}}=4.2$~\% for 1 ab$^{-1}$.
\end{abstract}

\section{Introduction}

The Standard Model predicts top quark production cross-section of 20~fb at 3 TeV.
Accurate determination of the cross-section as well as the forward-backward asymmetry allows to test the SM.  Deviations from the SM predictions could give indirect evidence of $Z'$ bosons, or, the measurements could be used to exclude the bosons to masses much higher than the $\sqrt{s}$ of the collider \cite{deCurtis,other}.  

The main background to be considered is $e^+ e^- \rightarrow W^+ W^-$ with gluon radiation.  This process is important as the cross-section is almost two orders of magnitude larger than that of the signal process.  The irreducible background  from $e^+ e^- \rightarrow bWbW$ has cross-section of 2.6~fb.

Pythia 6.1 was used to generate processes $e^+ e^- \rightarrow \gamma^*/ Z^0 \rightarrow q \bar{q}$ and $e^+e^- \rightarrow W^+ W^-$.  The detector response was simulated by Simdet 4.0 with CLIC settings, the beam spectrum was obtained using Calypso, and Hades was used for $\gamma \gamma$ background, integrating over four bunch crossings.

\section{ Event Reconstruction}


The $e^+ e^- \rightarrow t \bar{t}$ events have three possible topologies, which require different treatment in the reconstruction.  The top quarks are expected to decay $t \rightarrow b W$.  The $W$ boson can decay either to leptons, $W \rightarrow \ell \bar{\nu}$, or to quarks, $W \rightarrow q \bar{q}$.  The event is called hadronic (46~\%), if both $W$ bosons decay to quarks, leptonic (10~\%), if they both decay to leptons, and semileptonic (44~\%), if one $W$ decays to hadrons and the other to leptons.  Only those top quarks with the $W$ boson decaying to hadrons can be accurately reconstructed.

The main challenge in the reconstruction comes from jet clustering.  It is assumed that a hadronically decaying $t$ produces three jets and a leptonically decaying $t$ one jet and an isolated lepton.  Therefore the number of jets in the event is forced to be either six (hadronic case) or four (semileptonic case).  However, the angles between the produced quarks are rather small, typically of the order of $0.2$~rad.  Also especially the $b$ jets can be very soft, with energies less than typical gluon jet energies.  Therefore, the $b$ jet may be merged with one of the jets from $W$, or the $W$ jets merged together, with an additional gluon jet reconstructed.

The event reconstruction was started with lepton tagging.  The tagged leptons were required to have $E_\ell > 50$~GeV and no other tracks within 0.01~rad.  The efficiency to tag a lepton was assumed to be $90$~\%.
Jets were reconstructed using the JADE algorithm, requiring six jets when there were no tagged leptons in event and four jets otherwise.  Jet $b$-tagging was then performed, assuming 85 \% efficiency to identify a $b$ jet, 30 (10) \% efficiency to mistag a $c$ ($uds$) jet.  Only events with 1, 2 or 3 $b$-tagged jets were considered further.  

It was assumed that each top consists of one $b$-tagged jet and between one and three non-$b$-tagged jets.  Among the non-$b$-tagged jets, the combination that gave reconstructed mass closest to $m_W$, was searched.  When no leptons were tagged in the event, the combinations of jets with two reconstructed tops that minimised $|m_t-m(t_1)|+|m_t-m(t_2)|+|m_W-m(W_1)|+|m_W-m(W_2)|$ was selected.  In events with tagged leptons, only one top was reconstructed and thus the combination with the smallest $|m_t-m(t)|+|m_W-m(W)|$ was chosen. 


Reduction of the background was performed based on jet energies, number of tracks in each jet and the distances between the jets.  Typically $WW$ pair events have two very energetic jets and the rest of the jets are very soft, while in signal events all the jets tend to be in the mid-energy range.  Based on the distributions, for each event a probability was calculated for the event to be a signal event, $WW$ pair event or $e^+ e^- \rightarrow b \bar{b}$ event.  Also probabilities that the event is hadronic, leptonic or semileptonic signal event were obtained for use in the asymmetry measurement.

\section{Cross-section Measurement}

For the cross-section measurement, events were required to have more than $70$~\% probability to be a signal event if two or three of the jets were $b$-tagged, and more than $80$~\% probability, if only one jet was $b$-tagged.  Only events where the reconstructed masses satisfied $40$~GeV/$c^2 < m_W < 110$~GeV/$c^2$ and $115$~GeV/$c^2 < m_t < 240$~GeV/$c^2$ were selected.  The efficiency to select $e^+e^- \rightarrow t \bar{t}$ events and the fraction of these signal events in the selected sample after each selection are given in Table~1.  
\begin{figure}[hbt!]
\begin{center}
\label{tmas}
\epsfig{file=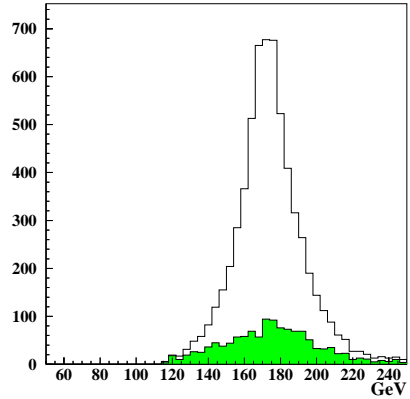, width=6 cm}\\
\caption{Reconstructed top mass distribution for the selected events.}
\end{center}
\end{figure}
The obtained efficiency of $24$~\% translates to $1.4$~\% accuracy in the cross-section determination.  The reconstruced top mass distribution is depicted in Figure~1.
\begin{center}
Table 1:  Signal event selection efficiency and purity for the cross-section measurement.\\ 
\begin{tabular}{|l|c|c|}
\hline
 & Efficiency (\%) & Purity (\%)\\
\hline
$b$-tagging, probability & 59 & 35 \\
\hline
$W$ mass, lepton & 25 & 74\\
\hline
$t$ mass & 24 & 80 \\
\hline
\end{tabular}
\end{center}

\section{ Forward-Backward Asymmetry}

For the forward-backward asymmetry measurement it is necessary to know the charge of the $t$ in each hemisphere.  In the case of semileptonic $e^+e^- \rightarrow t \bar{t}$ events this information is readily available from the decay of the leptonically decaying top, as the lepton charge depends on the $t$ charge: $ t \rightarrow W^+ \rightarrow \ell^+$ and  $ \bar{t} \rightarrow W^- \rightarrow \ell^-$.  Thus by identifying the charge of the lepton, the charge of the top is known.  Since the hadronic $t$ decay in the other hemisphere can be reconstructed, the value of $\cos{\Theta}$ can be obtained.  Using only semileptonic events limits the available statistics to $42$~\% of all available $e^+e^- \rightarrow t \bar{t}$ events.

Since the signal events are semileptonic $t \bar{t}$ events, background includes in addition to the $WW$ pair events and $b \bar{b}$ events the hadronic and leptonic $t \bar{t}$ events, and semileptonic events where the lepton does not originate from the $W$ decay.

The selected events were required to have exactly one tagged lepton.  Other selections depend on the number of $b$-tagged jets, as the dominating backgrounds are different.  In the case of only one $b$-tagged jet, the main background is $ W^+  W^-$ while with three $b$-tagged jets the dominating background are $ b \bar{b} $.  The properties of the leptons in these background processes are very different:  leptons from $W$ bosons typically have very high energies while leptons produced in semileptonic $b$ decays are less energetic.  The selection criteria included the probability of the event to be a $t \bar{t}$ event, the fraction of total reconstructed energy carried by the lepton and the cosine of the polar angle of the lepton.  In addition to the selection criteria listed in Table~2, it was required that the for the two probabilities $p(\mathrm{hadronic}) < (p(\mathrm{semileptonic})+0.06)$ to reduce the number of selected hadronic events.  Also, in the events where the was one $b$-tagged jet and altogether 6 jets, it was required that $p(WW)<0.12$ and $(p(WW)+p(b \bar{b}))<0.25$.  For the reconstructed masses, it was required that $40$~GeV/$c^2< m(W)< 120$~GeV/$c^2$ and $m(t) > 110$~GeV/$c^2$.
\begin{center}
Table 2: The selection criteria for the forward-backward asymmetry measurement.\\
\begin{tabular}{|l|c|c|c|}
\hline
$b$-tagged jets & 1 & 2 &3 \\
\hline
\hline
$t \bar{t}$ probability & $> 0.7$ &  $> 0.6$ & $> 0.6$ \\
\hline
$E_\ell / E_{tot}$ & $> 0.035$ &  $> 0.025$ & $> 0.025$ \\
\hline
$E_\ell / E_{tot}$ & $< 0.375$ &  $< 0.4$ & -  \\
\hline
$\cos{\Theta_\ell}$ & $> 0.05$ &  $> 0.05$ & -  \\
\hline
$\cos{\Theta_\ell}$ & $< 0.95$ &  - & -  \\
\hline
\end{tabular}
\end{center}

After imposing these selection criteria, semileptonic $t \bar{t}$ decays were selected with $19$~\% efficiency, and of the events in the sample $68$~\% were signal events.  Largest background was $WW$ pairs, $10$~\%, followed by leptonic events, $9$~\%.  Hadronic events were $6$~\% and semileptonic events with wrong sign lepton were $5$~\%.  Last contribution to the sample were $b \bar{b}$ events, 2~\%.  
\begin{figure}[hbt!]
\begin{center}
\epsfig{file=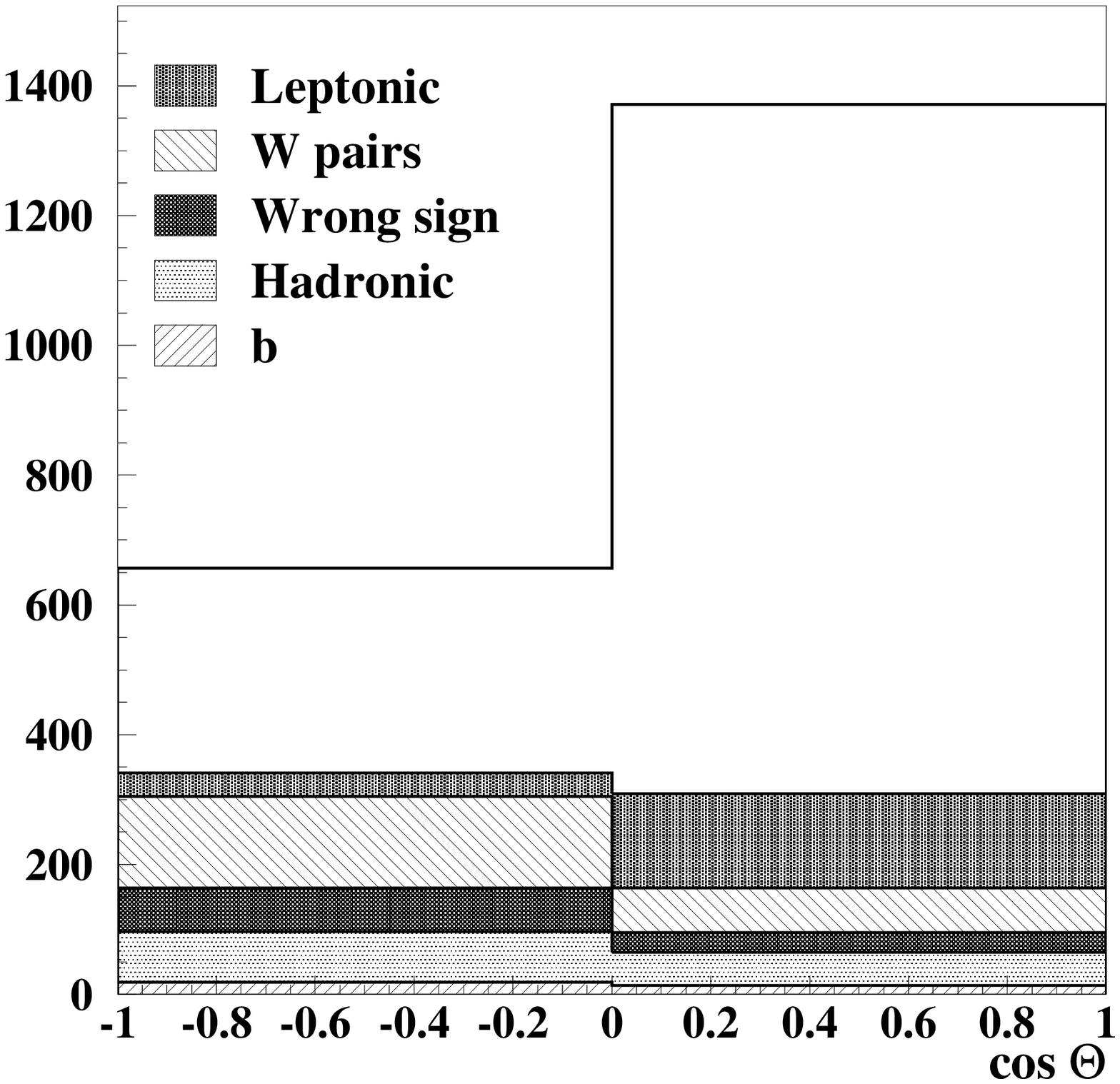,height=6.5cm}
\epsfig{file=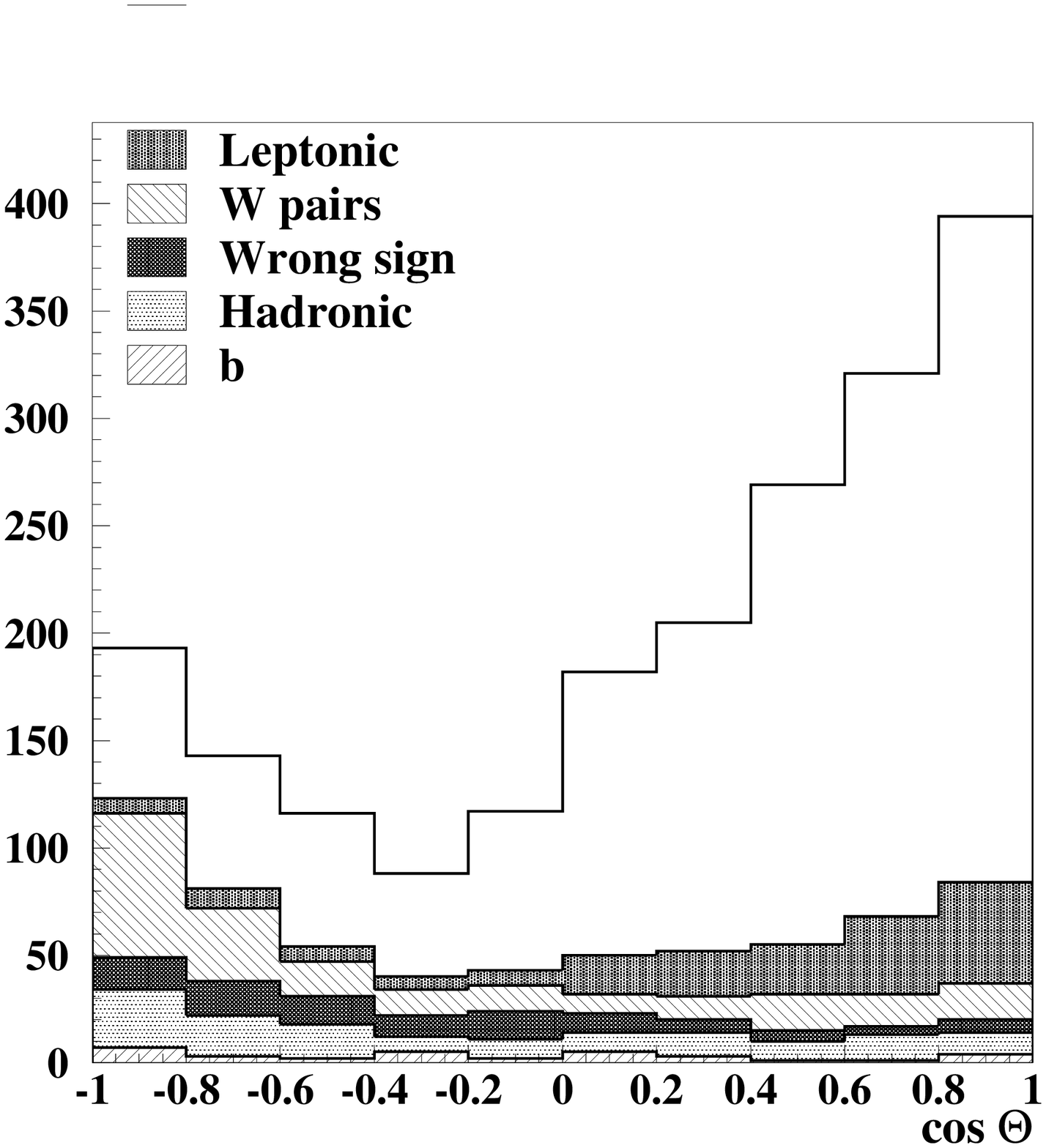,height=6.5cm}
\caption{The distributions for the study of forward-backward asymmetry.}
\end{center}
\end{figure}
Considering that the fully leptonic events can also be used for the asymmetry determination, the purity is $77$~\%.  With the $19$~\% efficiency to select semileptonic $t \bar{t}$ events and $11$~\% efficiency to select fully leptonic events, the value of asymmetry can be measured with $4.2$~\% accuracy with 1 ab$^{-1}$ of data, assuming perfect background subtraction.
The angular distributions with backgrounds are shown in Figure~2.

\section{ Conclusions}
A method for top reconstruction at CLIC was developed, and it was used to measure the total $e^+ e^- \rightarrow t \bar{t}$ cross-section $\sigma_{t \bar{t}}$ and the forward-backward asymmetry $A_{\mathrm{FB}}^{t \bar{t}}$.  For these measurements, background reduction was performed by using a probabilistic variable based on the jet energies, the number of tracks in each jet and the distance between the jets.  According to the analysis the accuracy of the total cross-section measurement will be of the order of ${\Delta \sigma_{t \bar{t}} }/{\sigma_{t \bar{t}}} = $ 0.014 and for the asymmetry  $ {\Delta A_{\mathrm{FB}}^{t \bar{t}}}/ {A_{\mathrm{FB}}^{t \bar{t}}} = $  0.042 for 1 ab$^{-1}$ of data.  As the asymmetry measurement is only targeted for semileptonic $t \bar{t}$ decays, the accuracy can be further improved by adding specific analysis for the fully leptonic channel and the fully hadronic channel.  The charge in the fully hadronic channel could be obtained by selecting events with a semileptonic $B$ decay in at least one hemisphere, or by using methods similar to the one used in the measurement of the $A_{\mathrm{FB}}^{b \bar{b}}$ \cite{other}. 

I am grateful to M. Battaglia and A. De Roeck for their comments and suggestions.

\end{document}